\documentclass[aps,pra,amsfonts,twocolumn,superscriptaddress,showpacs]{revtex4-1}

\usepackage{amsmath}
\usepackage{amsthm}
\usepackage{bbm}
\usepackage{bm}
\usepackage{braket}
\usepackage{color}
\usepackage{graphicx}

\def\E{{\cal E}}
\def\Emax{{\E}^{\max}_c}

\def\wt{\widetilde}
\def\rhomaxcan{\widetilde\rho^\mathrm{\,max}_c}
\renewcommand{\vec}[1]{\mathbf{#1}}

\DeclareMathOperator{\tr}{tr}

\newtheorem{theorem}{Theorem}
\newtheorem{conjecture}{Conjecture}

\begin{document}

\title{Quantum correlations of two-qubit states with one maximally mixed marginal}
\affiliation{Controlled Quantum Dynamics Theory, Department of Physics, Imperial College London, London SW7 2AZ, UK}
\affiliation{Mathematical Sciences, Brunel University, Uxbridge UB8 3PH, UK}
\author{Antony Milne}
\email{antony.milne@gmail.com}
\author{David Jennings}
\affiliation{Controlled Quantum Dynamics Theory, Department of Physics, Imperial College London, London SW7 2AZ, UK}
\author{Sania Jevtic}
\affiliation{Mathematical Sciences, Brunel University, Uxbridge UB8 3PH, UK}
\author{Terry Rudolph}
\affiliation{Controlled Quantum Dynamics Theory, Department of Physics, Imperial College London, London SW7 2AZ, UK}

\date{\today}
\begin{abstract}We investigate the entanglement, CHSH nonlocality, fully entangled fraction and symmetric extendibility of two-qubit states that have a single maximally mixed marginal. Within this set of states, the steering ellipsoid formalism has recently highlighted an interesting family of so-called `maximally obese' states. These are found to have extremal quantum correlation properties that are significant in the steering ellipsoid picture and for the study of two-qubit states in general.\end{abstract}

\pacs{03.67.Mn, 03.67.Bg}
\maketitle

\section{Introduction}

Quantum steering ellipsoids provide a faithful and intuitive representation of two-qubit states~\cite{Verstraete2002phd,Shi2011,Jevtic2013,Altepeter09}. If Alice and Bob each hold a qubit of a non-product state then Alice's Bloch vector is `steered' when Bob performs a local measurement. Given all possible measurements by Bob, the set of Bloch vectors to which Alice can be steered forms her steering ellipsoid $\E$ inside the Bloch sphere.  $\E$ is described by its centre $\vec c$ and a real, symmetric $3\times 3$ matrix $Q$. The eigenvalues of $Q$ give the squares of the ellipsoid semiaxes $s_i$ and the eigenvectors give the orientation of these axes. Not every $\E$ inside the Bloch sphere describes a physical two-qubit state; the necessary and sufficient conditions for physicality have recently been given in Ref.~\cite{Milne2014}.

In the steering ellipsoid formalism, the set of \emph{canonical} states is of particular importance. These correspond to two-qubit states in which Bob's marginal is maximally mixed. A general two-qubit state $\rho$ is transformed to its canonical state $\wt\rho$ by the local filtering operation~\cite{Shi2011}
\begin{eqnarray}
\label{eq:canonical}
\wt\rho&=&\left(\mathbbm{1}\otimes \frac{1}{\sqrt{ 2\rho_B}}\right)\,\rho\,\left(\mathbbm{1}\otimes  \frac{1}{\sqrt{ 2\rho_B}}\right)\nonumber\\
&=&\frac{1}{4}(\mathbbm{1}\otimes\mathbbm{1}+\vec{c} \cdot\bm{\sigma}\otimes\mathbbm{1}+\sum_{i,j=1}^3 \wt{T}_{ij}\,\sigma_i\otimes\sigma_j),
\end{eqnarray}
where $\rho_B=\tr_A \rho$. Since $\E$ is invariant under this transformation, only canonical states are needed to describe all possible physical steering ellipsoids. For a canonical state, Alice's Bloch vector coincides with the ellipsoid centre $\vec c$. The ellipsoid matrix of a general state $\rho$ is defined using its canonical state by $Q=\widetilde{T}\widetilde{T}^\mathrm{T}$. The ellipsoid semiaxes are therefore given by $s_i=|t_i|$, where $t_i$ are the signed singular values of $\wt{T}$. Without loss of generality we will say that the semiaxes are ordered such that $s_1\geq s_2 \geq s_3$. The chirality of $\E$ is defined as $\chi=\mathrm{sign}(\det \wt T)= \mathrm{sign} ({t}_1 {t}_2 {t}_3)$ and relates to the separability of the quantum state~\cite{Milne2014}; any entangled state must have $\chi=-1$.

\begin{figure}
\includegraphics[width=0.8\columnwidth]{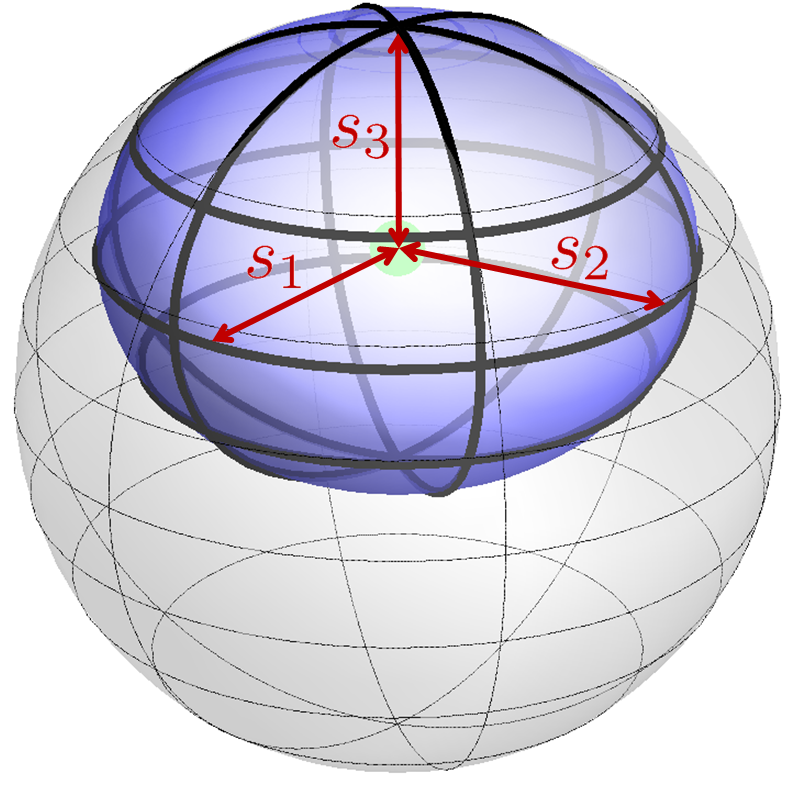}
\caption{\label{example_ellipsoid}An example $\Emax$ inside the Bloch sphere, with $c=0.4$. $\Emax$ has centre $\vec c=(0, 0, c)$ and semiaxes $s_1=s_2=\sqrt{1-c},s_3=1-c$.}
\end{figure}

In Ref.~\cite{Milne2014} we investigated extremal states lying on the physical-unphysical boundary by finding the largest volume physical $\E$ for any given $\vec c$. For $\vec c=(0, 0, c)$, the maximal volume ellipsoid $\Emax$ has major semiaxes $s_1=s_2=\sqrt{1-c}$ and minor semiaxis $s_3=1-c$ (see Fig. \ref{example_ellipsoid}). Since $\E$ is invariant under Bob's local filtering operations, the same $\Emax$ describes a whole manifold of states in which Bob's Bloch vector can take any value. However, by choosing the canonical state, which has Bob's marginal maximally mixed, we can associate with any given $\Emax$ a unique two-qubit state $\rhomaxcan$. This is the so-called `maximally obese' state, which forms a family parametrised by $0\leq c\leq 1$:
\begin{equation}\label{eq:max_state}
\rhomaxcan=\left(1-\frac{c}{2}\right)\ket{\psi_{c}}\bra{\psi_{c}}+\frac{c}{2}\ket{00}\bra{00},
\end{equation}
where $\ket{\psi_{c}}=\frac{1}{\sqrt{2-c}}(\ket{01}+\sqrt{1-c}\ket{10})$. Physically, $\rhomaxcan$ is Choi-isomorphic to the trace-preserving single-qubit amplitude-damping channel with decay probability $c$. With the exception of $c=1$, the maximally obese states are entangled; moreover, $\rhomaxcan$ is the state that maximises concurrence over the set of all two-qubit states that have steering ellipsoid centred at $\vec c$~\cite{Milne2014}.

Here we further investigate the set of canonical two-qubit states, with a particular focus on the family of maximally obese states. We find that $\rhomaxcan$ maximises three more measures of quantum correlation -- CHSH violation, fully entangled fraction, and negativity -- over the set of canonical two-qubit states with a given $\vec c$. We show that any maximally obese state must be either CHSH nonlocal or symmetrically extendible. Furthermore, in the context of steering ellipsoids, the entanglement properties of canonical states are found to correspond directly to simple geometric features of $\E$. Finally, we place necessary bounds on $\vec c$ for a general two-qubit state (i.e., one without any restriction on Bob's marginal) to be CHSH violating or useful for quantum teleportation.

\newpage
\section{CHSH violation and \\symmetric extendibility}

Consider the Clauser-Horne-Shimony-Holt (CHSH) scenario~\cite{Horne1969} with Alice and Bob sharing a canonical two-qubit state $\wt\rho$ of the form~\eqref{eq:canonical}. Alice can measure her qubit in one of the two directions $\bm{\alpha}$ or $\bm{\alpha}'$, and Bob can measure his qubit in $\bm{\beta}$ or $\bm{\beta}'$. Define the operator $B=\bm{\alpha} \cdot \bm{\sigma} \otimes (\bm{\beta}+\bm{\beta}')\cdot \bm{\sigma}+\bm{\alpha}'\cdot \bm{\sigma}\otimes (\bm{\beta}-\bm{\beta}')\cdot \bm{\sigma}$. The maximal CHSH violation gives a measure of Bell nonlocality and is given by $\beta(\wt\rho)=\max_{B}|\tr (\wt\rho B)|$, where the maximisation is performed over all directions $\bm{\alpha}, \bm{\alpha}', \bm{\beta}, \bm{\beta}'$. This gives~\cite{Horodecki1995}
\begin{equation}\label{eq:chsh}
\beta(\wt\rho)=2\sqrt{s_1^2 + s_2^2}.
\end{equation}

In the steering ellipsoid picture, the entanglement of a state depends on the centre vector $\vec c$, the size of $\E$ and its skew $\vec{c}^\mathrm{T}Q\vec{c}$~\cite{Jevtic2013}. In contrast to this, the CHSH nonlocality of a canonical state has a remarkably simple geometric interpretation: it depends on only the two longest semiaxes of $\E$ and not on the position or orientation of $\E$ inside the Bloch sphere.

\begin{theorem}\label{max_chsh}From the set of all canonical states with a given $\vec c$, the most CHSH nonlocal state is the maximally obese $\rhomaxcan$, as given in Eq. \eqref{eq:max_state}.
\begin{proof}
According to Eq. \eqref{eq:chsh}, we need to bound $s_1^2 + s_2^2$. The most CHSH nonlocal state will be entangled and so has $\chi=-1$. From the conditions for physicality given in Theorem 1 of Ref.~\cite{Milne2014} we have $s_1^2+ s_2^2\leq 1-c^2+2s_1 s_2 s_3-s_3^2$. As described in the appendix of Ref.~\cite{Milne2014}, we can use the Karush-Kuhn-Tucker conditions to show that the maximal volume $\Emax$ also maximises $2s_1 s_2 s_3-s_3^2$ for a given $\vec c=(0, 0, c)$. This $\Emax$ has $s_1=s_2=\sqrt{1-c}$ and $s_3=1-c$. We therefore see that $2s_1 s_2 s_3-s_3^2\leq (1-c)^2$, so that $s_1^2+ s_2^2\leq 1-c^2+(1-c)^2=2(1-c)$. This gives the bound $\beta(\wt\rho)\leq 2\sqrt{2(1-c)}$, which is met by $\rhomaxcan$.\end{proof}
\end{theorem}

Let us also consider the symmetric extendibility of maximally obese states. A bipartite quantum state $\rho_{AB}$ is \emph{symmetrically extendible} with respect to Alice if there exists a tripartite state $\rho_{AA'B}$ for which $\tr_A (\rho_{AA'B})=\tr_{A'} (\rho_{AA'B})$~\cite{Chen2014}. Originally introduced as a test for entanglement~\cite{Doherty2002}, symmetric extendibility has a number of operational interpretations. For example, a symmetrically extendible state cannot be used for one-way entanglement distillation~\cite{Nowakowski2009} or one-way secret key distillation~\cite{Myhr2009}. The relationship between symmetric extendibility and Bell nonlocality has also been studied; the results of Ref.~\cite{Terhal2003} show that a two-qubit state cannot be both symmetrically extendible and CHSH nonlocal. Although there exist (necessarily entangled) two-qubit states that are neither symmetrically extendible nor CHSH nonlocal, a maximally obese state must have one of these properties.

\begin{theorem}The family of maximally obese states $\rhomaxcan$ is partitioned into states that are symmetrically extendible and states that are CHSH nonlocal. $\rhomaxcan$ is symmetrically extendible for $1/2 \leq c\leq 1$ and CHSH nonlocal for $0\leq c< 1/2$.
\begin{proof}
The necessary and sufficient condition for a two-qubit state $\rho_{AB}$ to be symmetrically extendible with respect to Alice is $\tr (\rho_A^{\,2})\geq \tr(\rho_{AB}^{\,2})-4\sqrt{\det \rho_{AB}}$~\cite{Chen2014}. For $\rho_{AB}=\rhomaxcan$, as given in Eq. \eqref{eq:max_state}, we find that $\tr (\rho_A^{\,2})=(1+c^2)/2$, $\tr(\rho_{AB}^{\,2})=(2-2c+c^2)/2$ and $\det \rho_{AB}=0$. $\rhomaxcan$ is therefore symmetrically extendible if and only if $(1+c^2)/2\geq(2-2c+c^2)/2$, which gives $c\geq 1/2$.

The necessary and sufficient condition for a state $\rho_{AB}$ to be CHSH nonlocal is $\beta(\rho_{AB})>2$. From Theorem \ref{max_chsh}, we have that $\beta(\rhomaxcan)=2\sqrt{2(1-c)}$ and hence $\beta(\rhomaxcan)>2$ if and only if $c<1/2$.
\end{proof}
\end{theorem}

\section{Fully entangled fraction}

The fully entangled fraction of a bipartite state $\rho$ is defined by $f(\rho)=\max_{\ket{\phi}} \bra{\phi} \rho \ket{\phi}$, where the maximum is taken over all maximally entangled states $\ket{\phi}$~\cite{Bennett1996b}. $f(\rho)$ is an important quantity in entanglement distillation protocols~\cite{Bennett1996a} and relates directly to the fidelity of quantum teleportation~\cite{Horodecki1999a}.

For a canonical state $\wt\rho$ of the form \eqref{eq:canonical}, the fully entangled fraction is~\cite{Badziag2000}
\begin{equation}\label{eq:fully_ent_frac}
f(\widetilde\rho)=\frac{1}{4}(1+s_1+ s_2-\chi s_3),
\end{equation}
where we recall the ordering $s_1 \geq s_2 \geq s_3$. An entangled state must have $\chi=-1$; in this case $f(\wt\rho)$ depends only on the sum of the steering ellipsoid semiaxes $\sum_i s_i = \tr \sqrt{Q}$. Similar to CHSH nonlocality, the fully entangled fraction of a canonical state depends only on the size of $\E$ and not on its position or orientation.

\begin{theorem}\label{max_fidelity}From the set of all canonical states with a given $\vec c$, the state with the highest fully entangled fraction is the maximally obese $\rhomaxcan$, as given in Eq. \eqref{eq:max_state}.
\begin{proof}
According to Eq. \eqref{eq:fully_ent_frac}, we need to bound $s_1+s_2-\chi s_3$. Clearly we have $s_1+s_2-\chi s_3\leq s_1+s_2+s_3$. Since $s_3$ is the minor axis of a physical $\E$, we must have $s_3\leq 1-c$ in order for $\E$ to be inside the Bloch sphere, and so $s_1+ s_2+s_3\leq s_1+s_2+1-c$. From the Cauchy-Schwarz inequality, for any $n$-dimensional vector $\vec v$ we can bound the 1-norm $||v||_1$ and the 2-norm $||v||_2$ using $||v||_1\leq \sqrt{n} ||v||_2$~\cite{Hogben2007}. Applying this to $\vec v=(s_1, s_2)$, we have $s_1+s_2\leq \sqrt{2(s_1^2+ s_2^2)}$. From Theorem \ref{max_chsh}, $s_1^2+s_2^2\leq 2(1-c)$. We therefore see that $s_1+ s_2+s_3\leq 2\sqrt{1-c}+1-c$. This gives the bound $f(\widetilde \rho)\leq(1+\sqrt{1-c})^2/4$, which is met by $\rhomaxcan$.
\end{proof}
\end{theorem}

When Alice and Bob share a two-qubit state $\wt\rho$ to use as a resource for quantum teleportation, the average fidelity achieved is $F(\wt\rho)=[2f(\wt\rho)+1]/3$~\cite{Horodecki1999a}. Since $F(\wt\rho)$ increases monotonically with $f(\wt\rho)$, Theorem \ref{max_fidelity} shows that $\rhomaxcan$ is the optimal state to use for teleportation over all states that have Bob's marginal maximally mixed and Alice's Bloch vector equal to $\vec c$.

We can also consider this result in the Choi-isomorphic setting, using the fact that $\rhomaxcan$ is isomorphic to the amplitude-damping channel. Let us say that Alice prepares the Bell state $\ket{\psi^+}$ and sends one qubit of it to Bob through a trace-preserving quantum channel $\Phi$, intending the resulting shared state to act as a resource for teleportation. From the set of all non-unital maps $\Phi$ for which $\Phi(\mathbbm{1}/2)=(\mathbbm{1}+\vec c \cdot \bm \sigma)/2$ and $\vec c=(0, 0, c)$, the one that will maximise teleportation fidelity is the amplitude-damping channel.

These results complement previous studies of teleportation, which have shown that passing a resource state through a dissipative channel can enhance the average teleportation fidelity~\cite{Badziag2000,Ozdemir2007} as well as identifying the filtering operations that achieve optimal fidelity for a given resource state~\cite{OptimalTeleportation1,OptimalTeleportation2}.

\section{Concurrence and negativity}

We now consider two entanglement monotones, both of which range from 0 for a separable state to 1 for a maximally entangled state.

For a two-qubit state $\rho$, define the spin-flipped state as $\hat{\rho}=(\sigma_y \otimes \sigma_y)\rho^*(\sigma_y \otimes \sigma_y)$ and let $\lambda_1,...,\lambda_4$ be the square roots of the eigenvalues of $\rho \hat\rho$ in non-increasing order. The concurrence is then given by $C(\rho)=\max(0, \lambda_1-\lambda_2-\lambda_3-\lambda_4)$. Negativity is a measure for entanglement based on the Peres-Horodecki criterion~\cite{Horodecki1996a,Peres1996}. Let $\mu_\mathrm{min}$ be the smallest eigenvalue of $\rho^\mathrm{T_B}$; the negativity is then given by $N(\rho) = \max(0, -2\mu_\mathrm{min})$~\cite{Miranowicz2008,Vidal2002}.

In Ref.~\cite{Milne2014} we bounded the concurrence of any two-qubit state in terms of the volume of its steering ellipsoid. This gave us the bound $C(\wt\rho)\leq \sqrt{1-c}$ for a canonical state $\wt\rho$ of the form \eqref{eq:canonical}. The bound is saturated by maximally obese states $\rhomaxcan$. Our results on CHSH violation allow us to derive another result that is neither stronger nor weaker than this bound. Ref.~\cite{Verstraete2002} gives the bound $2\sqrt{2}C(\widetilde\rho)\leq \beta(\widetilde\rho)$. From Eq. \eqref{eq:chsh} and the ordering $s_1 \geq s_2 \geq s_3$ we then obtain $C(\wt\rho)\leq s_1$. Although this bound is distinct from $C(\wt\rho)\leq\sqrt{1-c}$, it is also saturated by maximally obese states.

Numerical results show that the negativity of a canonical state is bounded as $N(\wt\rho)\leq s_3$. As discussed in Theorem \ref{max_fidelity} we have $s_3\leq 1-c$, and so $N(\wt\rho)\leq 1-c$. Again, the bound is saturated by $\rhomaxcan$.

We therefore see that in the steering ellipsoid picture, the concurrence of a canonical state is upper bounded by the length of the major semiaxis while the negativity is upper bounded by the length of the minor semiaxis~\footnote{For any given steering ellipsoid $\E$, the state with highest concurrence is the canonical state~\cite{Milne2014}. The concurrence of \emph{any} two-qubit state is therefore upper bounded by the length of the major semiaxis.}. For maximally obese states, these entanglement measures are in fact equal to the lengths of these semiaxes and can thus be directly obtained from a geometric visualisation of $\Emax$.

As discussed in Ref.~\cite{Milne2014}, the maximally obese states form a special single-parameter class of the generalised Horodecki state (see, for example, Refs.~\cite{Horst2013,Miranowicz2008,Bartkiewicz2}). Other classes of the generalised Horodecki state have been studied before and were seen to have certain extremal properties. For example, the Verstraete-Verschelde states~\cite{Verstraete2002b} minimise the fully entangled fraction for a given concurrence and negativity; these states obey $C=[N+\sqrt{N(4+5N)}]/2$. Our maximally obese states $\rhomaxcan$ maximise concurrence for a given CHSH nonlocality and obey $C=\sqrt{N}$.

\section{Bounds for CHSH nonlocality and teleportation for general states}

\begin{figure}
\includegraphics[width=1\columnwidth]{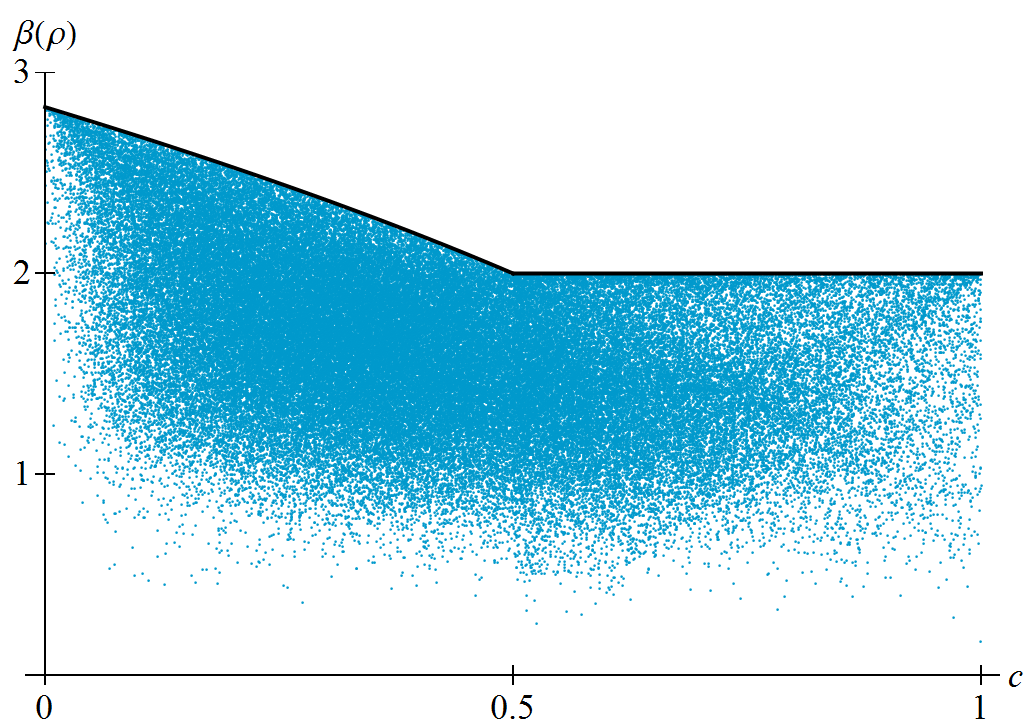}
\includegraphics[width=1\columnwidth]{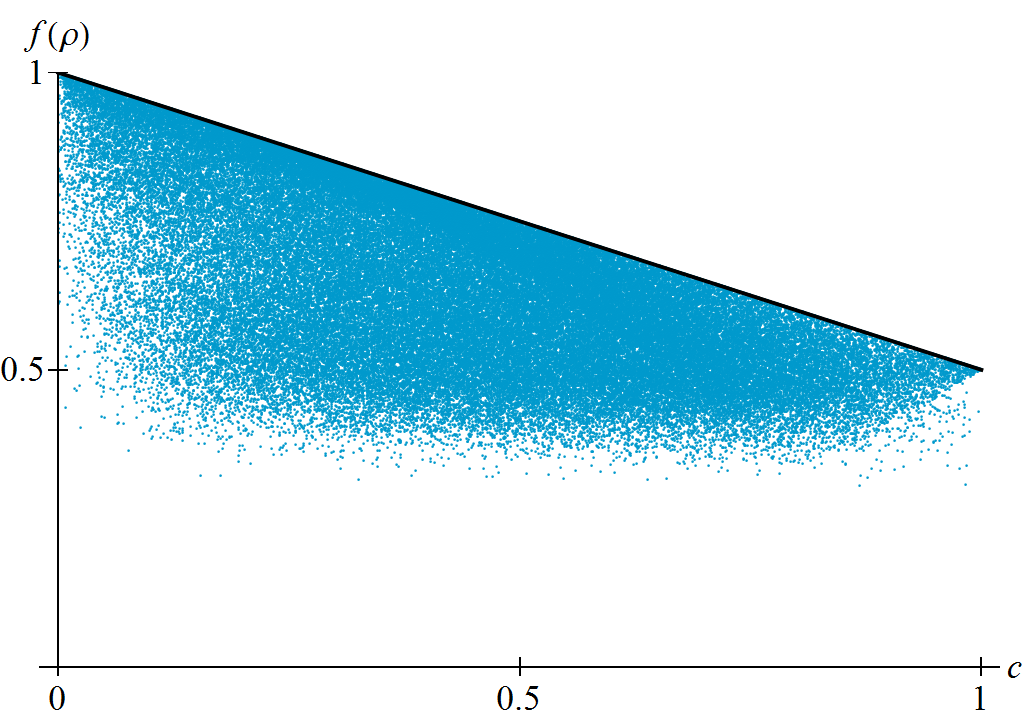}
\caption{\label{conjecture_graphs}The numerical evidence for Conjecture \ref{chsh_conjecture} (top) and Conjecture \ref{fidelity_conjecture} (bottom). Using $10^5$ random two-qubit states, we plot the CHSH nonlocality and fully entangled fraction against the magnitude of the steering ellipsoid centre. The conjectured bounds are shown as black lines.}
\end{figure}

CHSH nonlocality and fully entangled fraction are measures that do not transform straightforwardly under local filtering operations. The bounds given in Theorems \ref{max_chsh} and \ref{max_fidelity} for canonical states cannot therefore be used to analytically derive bounds for $\beta(\rho)$ and $f(\rho)$ for a general (i.e., not necessarily canonical) two-qubit state $\rho$. However, numerical investigations lead us to conjecture remarkably simple expressions for these bounds (see Fig. \ref{conjecture_graphs}).

\begin{conjecture}\label{chsh_conjecture}Let $\rho$ be a general two-qubit state with $\E$ centred at $\vec c$. The CHSH nonlocality is tightly bounded as $\beta(\rho)\leq\mathrm{max}[2\sqrt{2(1-c)},2]$.\end{conjecture}

This allows us to place a necessary bound on the steering ellipsoid for a general two-qubit state $\rho$ to be CHSH violating: to violate the CHSH inequality we need $\beta(\rho)>2$ and so $c<1/2$. We therefore see that a two-qubit state whose $\E$ is centred too close to the surface of the Bloch sphere cannot exhibit CHSH nonlocality.

\begin{conjecture}\label{fidelity_conjecture}Let $\rho$ be a general two-qubit state with $\E$ centred at $\vec c$. The fully entangled fraction is tightly bounded as $f(\rho)\leq 1-c/2$.\end{conjecture}

Recall that teleportation fidelity is related to fully entangled fraction by $F(\rho)=[2f(\rho)+1]/3$. Using only state estimation and classical communication, it is possible to achieve a teleportation fidelity of $2/3$~\cite{Horodecki1999a}. To beat this classical limit we require $f(\rho)>1/2$, and so we see that for all $c<1$ there exists $\E$ describing a state that achieves truly quantum teleportation. An optimal universal cloning machine achieves a fidelity of $5/6$~\cite{Bruss1998,Buzek1996}. To beat this limit we require $f(\rho)>3/4$ and hence $c<1/2$, which is the same bound as we obtained as a necessary condition for $\E$ to be CHSH violating.

\section{Outlook}

Steering ellipsoid centre $\vec c$ provides a natural parametrisation of two-qubit states and leads to geometric interpretations and simple bounds for several measures of quantum correlation. Whether these results can be easily extended to higher dimensional quantum systems remains to be seen. In particular, what would be the analogous family of maximally obese states in higher dimensions? It seems likely that the set of states Choi-isomorphic to higher dimensional amplitude-damping channels~\cite{HigherDimADC} will also have interesting maximal quantum correlation properties.

\begin{acknowledgments}
This work was supported by EPSRC. D.J. is funded by the Royal Society. T.R. would like to thank the Leverhulme Trust. S.J. acknowledges EPSRC Grant EP/K022512/1.
\end{acknowledgments}

\bibliography{references}

\end{document}